# A Survey of Quantum Programming Languages: History, Methods, and Tools

Donald A. Sofge, *Member, IEEE*

*Abstract*— Quantum computer programming is emerging as a new subject domain from multidisciplinary research in quantum computing, computer science, mathematics (especially quantum logic, lambda calculi, and linear logic), and engineering attempts to build the first non-trivial quantum computer. This paper briefly surveys the history, methods, and proposed tools for programming quantum computers circa late 2007. It is intended to provide an extensive but non-exhaustive look at work leading up to the current state-of-the-art in quantum computer programming. Further, it is an attempt to analyze the needed programming tools for quantum programmers, to use this analysis to predict the direction in which the field is moving, and to make recommendations for further development of quantum programming language tools.

## I. Introduction

THE importance of quantum computing has increased significantly in recent years due to the realization that we are rapidly approaching fundamental limits in shrinking the size of silicon-based integrated circuits (a trend over the past several decades successfully described by Moore's Law). This means that as we attempt to make integrated circuit components ever smaller (e.g. below 25nm in feature sizes), we will increasingly encounter quantum mechanical effects that interfere with the classical operation of the circuits.

Quantum computing offers a path forward that specifically takes advantage of quantum mechanical properties, such as superposition and entanglement, to achieve computational solutions to certain problems in less time (fewer computational cycles) than is possible using classical computing paradigms. Certain problems have been shown to be solvable exponentially faster on a quantum computer than has been achieved on a classical computer [1]. Furthermore, quantum parallelism allows certain functions that have thus far proven to be computationally intractable using classical computation to be executed in reasonable time (e.g. factoring large numbers using Shor's algorithm). Quantum algorithms may contain both classical and quantum components (as does Shor's algorithm), and can thus leverage the benefits of each.

However, existing classical (non-quantum) programming languages lack both the data structures and the operators necessary to easily represent and manipulate quantum data. Quantum computing possesses certain characteristics that distinguish it from classical computing such as the superposition of quantum bits, entanglement, destructive measurement, and the no-cloning theorem. These differences must be thoroughly understood and even exploited in the context of quantum programming if we are to truly realize the potential of quantum computing. We need native quantum computer programming languages that embrace the fundamental aspects of quantum computing, rather than forcing us to adapt and use classical programming languages and techniques as ill-fitting stand-ins to develop quantum computer algorithms and simulations. Ultimately, a successful quantum programming language will facilitate easier coding of new quantum algorithms to perform useful tasks, allow or provide a capability for simulation of quantum algorithms, and facilitate the execution of quantum program code on quantum computer hardware.

## II. Origins And History Of Quantum Computing

Prior surveys of quantum computing and quantum programming [2], [3] trace the origins of quantum computing and quantum programming to Feynman's 1982 proposal for constructing a quantum computer as a means of simulating other quantum systems, noting that a quantum computer may efficiently simulate a quantum system, whereas a classical computer simulation of a quantum system would require exponential resources both in memory space and computational time [4]. Preskill credits Paul Benioff [5] with making this proposal concomitantly with Feynman. However, the roots of quantum programming go far deeper than this, through the field of quantum information theory, in the work of Birkhoff and von Neumann on quantum logic in 1936 [6]. This work also formed the basis for quantum mechanics as it is currently practiced today.

Deutsch [7] investigated the computational power of quantum computers and proposed a quantum version of the Turing machine. He proposed one of the earliest quantum algorithms as a means of demonstrating the solution of a problem that would be difficult to solve using a classical computer yet would be quite easy to solve using a quantum algorithm.

Manuscript received September 12, 2007. This work was supported by the Naval Research Laboratory under NRL Work Order N0001406WX30002.

Donald A. Sofge is with the Navy Center for Applied Research in Artificial Intelligence, Naval Research Laboratory, Washington D.C. (phone: 202-404-4944; fax: 202-767-2166; e-mail: Donald.Sofge@nrl.navy.mil).

The difficulty of solving certain problems using computers is often characterized by analogy with hypothetical Turing machine based solutions [8]. Alan Turing showed that the capability of any general purpose computer could be simulated with a Turing machine. By studying the time (# computational cycles) and space (memory requirement) complexity of problems, we can classify problems according to their Turing complexity (e.g. deterministic polynomial (P), non-deterministic polynomial (NP), exponential (EXP)). The Deutsch-Jozsa algorithm [1] was designed in 1992 to maximally illustrate the computational advantage of quantum computing over classical computing. The algorithm determines a property of a binary input string (either *constant* or *balanced*) in a single step, whereas determining this property on a classical computer grows exponentially in computational complexity with the length of the bit string. Unfortunately, this result has not yet resulted in the creation of any particularly useful quantum algorithms for typical data processing needs.

Since that time, substantial effort has been made to characterize the complexity of problems using quantum Turing machine models versus classical Turing machine models. While there is still some question regarding certain complexity equivalence classes (e.g. does BQP=NP?), there is widespread agreement amongst researchers in this area that quantum computing has not yet been proved to generally move problems from one complexity class under classical computing to a lower complexity class under quantum computing [9]. However, this does not mean that nontrivial quantum computers once built will not allow substantial computational gains for solving certain problems over currently known techniques using classical computers. An important thread in quantum computing research is to define problem classes and applications that result in significant speed-ups using quantum instead of classical computers.

The invention of linear logic by Girard in 1987 [10] has also played a significant role in the formulation of recent quantum programming languages, specifically those based upon lambda calculus (described in section III.b). Linear logic differs from classical logic in that assumptions (states, or inputs) and hypotheses are considered resources that may be consumed. It provides a means for resource control. Linear logic differs from usual logics such as classical or intuitionistic logic where the governing judgement is of truth, which may be freely used as many times as necessary.

For example, suppose *A* represents water, *B* represents cold (or a freezing process), and *C* represents ice. Then $A \wedge B \rightarrow C$. But after the process is applied, resource *A* is consumed and is no longer available. Research in using linear logic is still quite active, and linear logic is important element to many current quantum programming language development efforts.

The first practical steps toward formulating a quantum programming language were made by Knill in 1996 in his proposal for conventions for a quantum pseudocode [11], and his description of the quantum random access machine (QRAM) model of a quantum computer. The QRAM model is built upon the (probably accurate) assumption that any practical quantum computer will in fact be a classical machine with access to quantum computing components, such as qubit registers. QRAM defines a set of specific operations to be performed on computer hardware including preparation of quantum states (from classical states), certain unitary operations, and measurement. Knill's quantum pseudocode provides a syntax for describing qubits, qubit registers, and operations involving both classical and quantum information. While extremely useful, Knill's proposal falls short of possessing all of the needed characteristics of a real quantum programming language due to its informal structure, lack of strong typing, and representation of only some of the quantum mechanical properties needed.

A variety of tools have been created for simulating quantum circuits and modest quantum algorithms on classical computers using well-known languages such as C, C++, Java, and rapid prototyping languages such as Maple, Mathematica, and Matlab. A good on-line reference for these simulators is

*http://www.quantiki.org/wiki/index.php/List_of_QC_simulators*

While simulators may provide an excellent means for quickly learning some of the basics concepts of quantum computing, they are not substitutes for actual quantum programming languages since they are designed to run only on classical computer architectures, and will not realize any of the computational advantages of quantum computing.

### III. A Taxonomy of Quantum Programming Languages

Quantum programming languages may be taxonomically divided into (A) imperative quantum programming languages, (B) functional quantum programming languages, and (C) others (may include mathematical formalisms not intended for computer execution). In addition, Glendinning [12] maintains an online catalog of quantum programming languages, simulation systems, and other tools. Finally, a slightly more dated but at the time quite comprehensive and still useful survey of quantum computer simulators was provided by Julia Wallace in 1999 [13].

Early quantum programming language development efforts focused on exploring the Quantum Turing Machine (QTM) model as proposed by Deutsch [7]. While interesting and informative from the standpoint of understanding computational complexity of problem classes with respect to quantum computing, it did not result in practical tools for programming quantum computers.

The quantum circuit model quickly became the driving force in quantum programming. The Deutsch-Jozsa algorithm, Quantum Fourier Transform, Shor's factoring algorithm, and Grover's algorithm were all described using the quantum circuit model [8]. In order to build this into a language (instead of just designing circuits), Knill [11] proposed a quantum programming pseudocode that, along with adapted imperative programming languages such as C and C++, resulted in the first imperative quantum programming languages (e.g., QCL). These languages built upon the QRAM model of quantum computation, assumed classical flow control with both classical and quantum data, and allowed interleaved measurements and quantum operations.

The use of imperative quantum programming languages gave way to a multitude of functional quantum programming languages such as QFC, QPL, and QML (described below), mostly based upon the QRAM model, but also increasingly utilizing the work in mathematical logic to define better operational semantics for quantum computing, and also including the tool of linear logic discussed previously.

### A. Imperative Programming Languages

*Imperative* programming languages, also known as *procedural* languages, are fundamentally built upon the use of statements to change the global state of a program or system of variables. Common classical imperative languages include FORTRAN, Pascal, C, and Java. These may be contrasted with *functional* (or *declarative*) languages such as Lisp, APL, Haskell, J, and Scheme, in which computation is based upon the execution of mathematical functions.

Imperative quantum programming languages today are largely descendents of Knill's proposed quantum pseudocode and the QRAM model of quantum computing. Arguably the first "real" quantum programming language [2] was QCL, developed and refined from 1998-2003 by Bernhard Ömer [14]-[18]. QCL (Quantum Computation Language) utilizes a syntax derived from C, and also provides a full quantum simulator for code development and testing on a classical computing platform. In support of the classical+quantum model of computation as envisioned with the QRAM virtual hardware model, QCL also provides a full classical programming sublanguage. High-level quantum programming features include automatic memory management, user defined operators and functions, and computation of the inverse of a user defined operator. QCL may be downloaded from the web at *http://tph.tuwien.ac.at/~oemer/qcl.html*

Betelli et al. [19] devised an imperative language based upon C++. The language was created in the form of a C++ library, and thus can be compiled. It is also maintained as a downloadable source code through *http://sra.itc.it/people/serafini/qlang/*

Important features of this language include construction and optimization of quantum operators at run-time, classes for basic quantum operations like QHadamard, QFourier, QNot, QSwap, and Qop. The language also supports user definition and construction of new operators. Another important feature offered by this language is simulation of the noise parameters in the simulator.

This language is occasionally called Q Language by the authors, but another perhaps more prominent functional language exists with the same name, (for which the Q stands for eQuation), thus use of this name should probably be avoided in at least one of the two contexts.

Another imperative quantum programming language, called qGCL (quantum Guarded Command Language), was proposed by Sanders & Zuliani [20] as a derivative of Dijkstra's guarded command language, intended more for algorithm derivation and verification rather than programming.

Most recently (October 2007) Mlnarik [21] introduced the imperative quantum programming language LanQ which uses C-like syntax and supports both classical and quantum process operations, including process creation and interprocess communication. Mlnarik provides full formalized syntax for the language, operational semantics, proves type soundness (eliminating type errors) for the non-communicating part of the language, and can be used for proving correctness of implemented quantum algorithms. The author also provides a publicly accessible simulator for LanQ at

*http://lanq.sorceforge.net/*

### B. Functional Quantum Programming Languages

*Functional* (or *declarative*) programming languages do not rely upon the update of a global system state, but instead perform mathematical transformations by executing mappings from inputs to outputs. Most recent developments in quantum programming have focused on the use of functional rather than imperative languages. The languages in this category are based upon the concept of a lambda calculus. Lambda calculi are constructions from mathematical logic used to investigate the properties of functions, such as computability, recursion, and stopping.

Lambda calculi may be considered the smallest universal programming languages. They consist of a single transformation rule (variable substitution) and a single function definition scheme. They are universal in the sense that any computable function can be expressed and evaluated using this formalism. It is thus equivalent to the Turing machine formalism. However, lambda calculi emphasize the use of transformation rules, and do not care about the actual machine implementing them.

Lambda calculi were first proposed by Alonzo Church and Stephen Cole Kleene in the 1930s, and used by Church in 1936 [22], [23] to address the decision problem (*entscheidungs-problem*) challenge proposed by David Hilbert. Lambda calculi can be used to define what a computable function is. The question of whether two lambda calculus expressions are equivalent cannot be solved by a general algorithm. This was the first question, even before the halting problem, for which undecidability could be proved. Since their invention, lambda calculi have greatly influenced classical functional programming languages such as Lisp, ML and Haskell.

In 1996 Maymin [24] proposed a quantum lambda calculus to investigate the Turing computability of quantum algorithms. While Maymin's lambda calculus was found to efficiently solve NP-complete problems, it was unfortunately found to be more expressive than any physically realizable quantum computer [25].

In 2004 van Tonder [26] defined a quantum lambda calculus for pure quantum computation (no measurements take place), analyzed the non-duplicability of quantum states through use of linear logic, and argued that the language has the same equivalent computational capabilities as a quantum Turing machine. Quantum algorithms are implemented in a quantum simulator built upon the Scheme programming language.

Also in 2004 Selinger [27] proposed a functional quantum programming language called QFC (Quantum Flow Charts) which represents programs via functional flow charts, and an equivalent form which utilizes textual syntax called QPL (Quantum Programming Language). These languages rely upon the notion of using classical control and quantum data, and build upon a lambda calculus model to handle both classical and quantum data within the same formalism. These languages may be compiled using the QRAM virtual quantum computer model. However, they still lack many desirable aspects including higher-order features and side-effects.

In 2004 Danos [28] studied a one-way (non-reversible) model of quantum computation that included notation for entanglement, measurement, and local corrections.

In 2005 Perdrix [29] defined a type system that reflects entanglement of quantum states. This too was based upon a lambda calculus. Altenkirch and Grattage [30], [31] developed a functional quantum programming language called QML in which control as well as data may be quantum. QML is based upon a linear logic [10] (described previously in Section II), but focuses on the elimination of weakening (discarding a quantum state) instead of contraction (duplication of quantum state). Several researchers have proposed domain-specific functional quantum programming languages implemented in Haskell, following in the style of Selinger's QPL, using linear logic and lambda calculi, and building upon the 2001 work of Mu and Bird [32] in which quantum programming is modeled in Haskell through definition of a data type for quantum registers. Sabry [33] extended this model to include representation of entangled states. Other related efforts in this vein include the work of Vizzotta and da Rocha Costa [34], Karczmarczuk [35], and Skibinski [36].

*C. Other Quantum Programming Language Paradigms*

A substantially different approach to quantum programming was offered by Freedman, Kitaev, and Wong [37] based upon the simulation of topological quantum field theories (TQFT's) by quantum computers. TQFT's provide a more robust model of quantum computation by representing quantum states as physical systems resistant to perturbations. Quantum operations are determined by global topological properties, such as paths that particles follow. This radically different approach to quantum computing may provide new insights and lead to the creation of new types of quantum algorithms. However, as currently formulated it only deals with evolution of state and does not include a measurement process.

A number of efforts have been made in recent years to define languages to support quantum cryptographic protocols, and specifically focus on the inclusion of communication between quantum processes. Such processes may be local or nonlocal, thus giving rise to distributed quantum programming specifications. A prime example of this is Maurer's [38] specification of the cQPL language, based upon Selinger's QPL, but with extensions added to support communication between distributed processes.

Another thread of development has focused on the use of quantum process algebras, such as QPAlg (quantum process algebra) by Jorrand and Lalire [39] to describe interactions between classical and quantum processes. Gay and Nagarajan [40] define the language CQP (communicating quantum processes) for modeling systems combining classical and quantum communication, with particular emphasis on applications such as quantum cryptography. CQP is designed specifically to provide complete protocol analysis, prove type soundness, and lead to methods for formal verification of systems modeled in the language. Both QPAlg and CQP influenced the design of the imperative programming language LanQ described previously.

Adão and Mateus [41] give a process calculus for security protocols built upon the QRAM computational model with an added cost model. Udrescu et al. [42] describe a hardware description language for designing quantum circuits similar to those used for VLSI design.

The use of Girard's linear logic and lambda calculus has spurred a considerable amount of recent work in the formulation of mathematical formalisms for quantum computing that include such semantic features as entanglement, communication, teleportation, partial and mixed quantum states, and destructive measurement. While exciting and worthwhile, these formalisms generally fall short of actually specifying specific programming languages with full syntax, well defined operators, and simulators and compilers needed to implement Shor's or Grover's algorithms or to create and test new quantum algorithms. These efforts do help to address shortcomings in quantum theory itself as discussed in the next section.

IV. CHALLENGES IN QUANTUM PROGRAMMING LANGUAGE DEVELOPMENT

The difficulties in formulating useful, effective, and in some sense universally capable quantum programming languages arise from several root causes. First, quantum mechanics itself (and by extension quantum information theory) is incomplete. Specifically missing is a theory of measurement. Quantum theory is quite successful in describing the evolution of quantum states, and even in predicting probabilistic outcomes after measurements have been made, but the process of state collapse is (with a few exceptional cases) not covered. So issues such as decoherence, diffusion, entanglement between particles (or entangled state, of whatever physical instantiation), and communication (including teleportation) are not well defined from a quantum information (and by extension quantum computation) perspective. Work with semantic formalisms and linear logic attempt to redress this by providing a firmer basis in a more complete logic consistent with quantum mechanics. These logic-based formalisms (once validated) may then be combined with language syntax and other programming language features to more accurately and completely reflect the potential capabilities of quantum computing.

A second key source of difficulty is the lack of quantum computing hardware for running quantum algorithms. Given no specific set of quantum operations (e.g. specific quantum gates, ways that qubits are placed in superposition or entangled), then no guidance is available to computer scientists designing quantum programming languages as to what data structures should be implemented, what types of operations to allow, what features to disallow (such as abiding by the no-cloning theorem), and how to best design the language to do what quantum computing does best (e.g., is an imperative language like C or Java a better place to start, or a functional language like LISP or APL?). A third source of difficulty is the paucity of practical applications for quantum computing. Shor's algorithm generated tremendous excitement over the potential of quantum computing in large part because most educated individuals could immediately recognize the value of the contribution, at least in the context of code breaking. Grover's algorithm also provides a speed-up over classical computation for a specific class of problems (searching unsorted lists), from $O(N)$ to $O(N^{1/2})$, which is still quite impressive, especially for large N. Shor's algorithm is capable of achieving its impressive speedup by exploiting the structure of the problem in the quantum domain, recasting the factoring problem as a period-finding problem which may then be solved in parallel. Grover's algorithm is arguably more general, and variants of Grover's algorithm have been proposed for a variety of problems [43], [44].

Finally, there is much on-going research to understand the quantum physics behind quantum information, and by extension quantum computing. We have previously discussed several quantum computing conceptual models including QRAM, the quantum Turing machine, and the quantum circuit model. However, recent work by in the area of adiabatic quantum computing by Dorit Aharonov [45] and weak measurements by Jeff Tollaksen [46] and Yakir Aharonov [47] arguably falls under none of these categories, yet these methods represent potentially exciting new directions for realizing quantum computing. But this wealth of models also serves to further complicate the process of establishing a set of practical tools for writing quantum programs, namely quantum programming languages.

## V. A Path Forward for Quantum Programming Language Development

Julia Wallace [13] in her quite thorough survey of quantum simulators notes that most of the simulators and even some of the languages developed could only really implement one algorithm, usually Shor's (some could do Grover's as well, and some could do more). Selinger [2] comments *"each new algorithm seems to rely on a unique set of 'tricks' to achieve its goal"*. The hope, of course, is that one can take a set of known quantum algorithms, such as Shor's, Grover's, Deutsch-Jozsa, and a few others, and from this set to grow a methodology for quantum programming, or at least generate a few new useful quantum algorithms. Unfortunately, this set of examples is too small and specialized to grow a methodology for quantum programming, and generating useful new quantum algorithms has proven quite difficult.

The work in operational semantics, lambda calculi, and linear logic (described above) and the inclusion of a complete programming grammar and syntax will undoubtedly yield multiple powerful languages for developing quantum algorithms. These languages will probably be of the functional class, but they may take other forms as well. However, the question of what sorts of quantum programs will be written and what applications they will address is still open, and the answer to this question will serve as a major driving factor in shaping the quantum programming languages to come. This is supported analogically by comparison with the numerous application domains for classical computing and the even larger number of classical programming languages created to develop products from them in the form of application programs.

Perhaps the greatest potential for useful computational gains of quantum computing over classical computing may be found with the class of problems known hidden subgroup problems [48], [49]. The factoring problem that Shor's algorithm addresses, as reformulated by Shor, falls within this class. One strategy, then, would be to try to understand which problems may be mapped into the hidden subgroup problem class, and then investigate those problems for development of quantum algorithms to address them.

Finally, the realization of quantum computing hardware will significantly drive the design of quantum language operators and data structures for quantum programs to be run on that hardware. We can already make some guesses about early hardware (e.g. superposition of a small number of locally contiguous qubits, preparation of states using Hadamard gates, rapid decoherence of quantum states). But if qubits are represented by photons, such guesses may very well be wrong.

When considering the need for quantum algorithms, and asking what kinds of quantum algorithms are needed, it is useful to think of a (classical) programmer as possessing a toolbox filled with a rich diversity of tools (algorithms) from which he/she can construct solutions (new programs) to client problems. From this basis we can proceed cautiously to design small but interesting quantum algorithms, gradually expanding the universe of basic quantum functions and capabilities to put into our quantum programming toolbox.